\documentclass{aa}
\usepackage{graphicx}

\begin{document}
  \title{Disk or Halo white dwarfs?}
\subtitle{Kinematic analysis of high proper motion surveys}
   \author{A. Spagna,
    \inst{1}
    D. Carollo,
    \inst{1}
    M.G. Lattanzi,
      \inst{1}
        B.\ Bucciarelli
    \inst{1}
        }

 \institute
   {INAF, Osservatorio Astronomico di
   Torino, I-10025 Pino Torinese, Italy
    }

   \date{Received 13 April 2004; accepted 21 July 2004}

   \abstract{
We present an alternative method for the kinematic analysis of
high proper motion surveys and discuss its application to the
survey of Oppenheimer et al.\ (2001) for the selection of reliable
halo white dwarfs (WDs).
The local WD space density we estimate  is $\rho_{\rm WD} \simeq 1
\div 2 \cdot 10^{-5}$ M$_{\odot}$pc$^{-3}$, which is about an
order of magnitude smaller than the value derived in Oppenheimer
et al.\ (2001), and is consistent with the values obtained from
recent reanalyses of the same data (e.g.\ Reid et al.\ 2001,
Reyl\'e et al.\ 2001, Torres et al.\ 2002, Salim et al.\ 2004).
 Our result, which corresponds to a fraction of 0.1\% $\div$ 0.2\%
of the local dark matter, does not support the scenario suggested
by the microlensing experiments that ancient cool WDs could
contribute significantly to the dark halo of the Milky Way.
      }
   \maketitle
%

\section{Introduction}
One of the most recent challenges in observational astronomy is to
explain the nature of the objects that produced the microlensing
events towards the Magellanic Clouds (Alcock et al.\ 2000). The
most obvious candidates for these events are ancient white dwarfs,
so that several projects have been carried out in recent years to
reveal the existence of such hidden population of dim sources (see
Hansen \& Liebert 2003 for a review). The most extensive survey to
date is that of Oppenheimer et al.\ (2001, OHDHS). They discovered
38 suspected halo white dwarfs and derived a local density of
$\sim 1.1 \cdot 10^{-4}$ M$_\odot$ pc$^{-3}$, which corresponds to
a fraction of 1-2\% of the halo dark matter in the vicinity of the
Sun. Different authors challenged these results on the basis of
the age estimates of the candidates (Hansen 2001, Bergeron 2003),
or after a reanalysis of the kinematic data (e.g.\ Reid, Sahu \&
Hawley\ 2001; Reyl\'e et al.\ 2001, Flynn et al.\ 2003, Torres et
al.\ 2002).

In any event, all those studies evidence a significant
contamination of thick disk objects affecting the halo WD sample,
and point out the basic problem of defining an accurate procedure
to deconvolve the halo and thick disk populations on the basis of
their kinematic and photometric properties.

In this paper we describe a general statistical method designed to
reject objects with disk kinematics and isolate probable halo
members from the screening of kinematically selected samples.
Finally, we discuss the results obtained with this method when
applied to the OHDHS survey, and compare them to the preliminary
results derived from the GSC II-based new high proper motion
survey in the Northern hemisphere by Carollo et al.\ (2004).


\section{SSS Halo WD survey}
The OHDHS survey was based on digitized, photographic Schmidt
plates (R59F and B$_J$ passbands) from the SuperCOSMOS Sky Survey
(SSS, Hambly et al. 2001). They analyzed 196 three epoch plates
(IIIaJ, IIIaF and IV-N) covering an area of 4165 square degrees
near the South Galactic Pole (SGP). The  magnitude limit of the
survey is of R59F = 19.8, while the proper motion limits are
0.33$\arcsec $yr$^{-1} < \mu < 10\arcsec $yr$^{-1}$.
They found 98 WDs, whose tangential velocities were derived from
the measured proper motions and photometric distances estimated
via a linear color magnitude (CM) relation, $M_{B_J}$ vs.\
$B_J-R59F$, calibrated by means of the WD sample with available
trigonometric parallaxes published by Bergeron, Ruiz \& Legget
(1997).
 The kinematic analysis of this sample was made in the two dimensional
 (U,V) plane, after assuming that the third galactic velocity component was zero ($W=0$).
Thick disk contaminants were rejected with a 2$\sigma$ threshold,
$\sqrt{U^2+(V+35)^2}>95$ km $s^{-1}$, which would correspond to a
86\% confidence level in the case of a non-kinematically selected
sample. In this way, 38 WDs were considered as halo members, from
which a space density of $\rho_{\rm WD}\simeq 1.1\, 10^{-4}$
M$_\odot$ pc$^{-3}$ was computed, assuming 0.6 M$_\odot$ for the
average WD mass.

As mentioned in the previous section, these results were
critically revised by several authors. In particular, an
independent kinematic analysis of the OHDHS sample was performed
by Reid et al.\ (2001), who noted that the resulting distribution
of the WDs in the (U,V) diagram seems more compatible with the
high velocity tails of the thick disk. They computed $(U,V)$
components assuming that the unknown radial velocity is null
($V_{r} = 0$) and selected halo WDs with the crude but robust
criterion of accepting objects with retrograde motion only (4
objects).  This leads to a more conservative value of the density,
$\rho_{\rm WD}\simeq 1.8\, 10^{-5}$ M$_\odot$ pc$^{-3}$.

Recently, Salim et al.\ (2004) reanalyzed the WD sample of OHDHS
on the basis of new spectroscopic and photometric measurements.
Radial velocities of 13 WDs with H$_{\alpha}$ line, and standard
Johnson-Cousins photometry for half of the sample were obtained.
In addition, distances were redetermined with the CCD photometry
by means of the theoretical color magnitude relation for hydrogen
and helium atmospheres published by Bergeron, Leggett \& Ruiz
(2001). Salim et al.\ (2003) confirmed the results of OHDHS with
the same 95 km s$^{-1}$ (2$\sigma$) threshold, but showed that a
minimum density, $n_{\rm WD}\simeq 3.1\, 10^{-5}$ pc$^{-3}$ is
attained with a higher, more conservative, threshold of 190 km
s$^{-1}$.



\section{Kinematic analysis} \label{Sect:kinematics}
 The kinematic analysis
of the WD sample drawn from a proper motion limited survey,
including the choice of a criterium for rejecting the contaminant
disk WDs and select the true halo WDs, is one of the critical
steps of this kind of studies.

As the velocity distribution of the disk(s) and halo population do
partially overlap (Fig.~1), it is not possible to infer
univocally, on the basis of kinematic data alone, the parent
population of every object. Nevertheless, it is always possible to
test if an object is, or is not, consistent with the velocity
distribution of a certain population once a value for the
confidence level is chosen.

Here, we retain  as halo WDs those objects whose kinematics is
{\it not} consistent with the velocity distribution of the thick
disk population\footnote{Implicitly, we assume that besides the
thick disk WDs, this criterion rejects the ``slowest'' thin disk
objects  as well.} given a certain confidence level; this allows
the identification of halo WDs while limiting the contamination of
high velocity thick disk objects.
 Unless corrected for the incompleteness due to the fraction of rejected halo WDs
whose kinematics is compatible with that of the thick disk
population, it is clear that this procedure can only provide a
lower limit to the actual density.

An alternative, and potentially more rigorous procedure, is a
Maximum-likelihood analysis that fits simultaneously the
superposition of two or more populations (see e.g.\ Nelson et al.\
2002, Koopmans \& Blandford 2002). In this case however, because
of the small size of the samples, further assumptions on the
kinematics and the formation process (IMF, age, etc.) of {\it all}
the populations involved are usually necessary.

\subsection{Schwarzschild distribution}
We assume that the probability that the galactic velocity
components (U,V,W) of an object in the solar neighborhood
belonging to a certain stellar population lies in the element of
velocity space $d^{3}\bar{v}=dU dV dW$ is well described by a
Schwarzschild distribution:
\begin{equation}
 p(\bar{v})=\frac {1}    {(2\pi)^{3/2}\sigma_{U}\sigma_{V}\sigma_{W}}
\exp \left[ -\frac{U^2}{2\sigma_U^2} - \frac{(V-V_{0})^2}
{2\sigma_V^2} -
 \frac{W^2}{2\sigma_W^2} \right]
 \label{eq:UVWdistribution}
\end{equation}

which represents a trivariate gaussian ellipsoid, where $V_0$
indicates the rotation lag with respect to the LSR and $\sigma_U$,
$\sigma_V$, and $\sigma_W$ the velocity dispersions.

In practice, the galactic components need to be derived from the
observed tangential and radial velocity components $(V_\alpha,
V_\delta, V_r)$:
\begin{equation}
 \left[
 \begin{array}{c}
    U \\ V \\ W \\
  \end{array}
  \right]
  = {\bf G}_{2000}
 \left[
 \begin{array}{l}
   V_\alpha\\
   V_\delta  \\
    V_r \\
  \end{array}
  \right]
  +  \left[
 \begin{array}{l}
   U_\odot\\
   V_\odot  \\
    W_\odot \\
  \end{array}
  \right]
 \label{UVW}
\end{equation}
 where {\bf G}$_{2000}={\bf G}(\alpha,\delta)$ is the
transformation matrix from the equatorial coordinates system
(J2000) to the galactic system, which depends explicitly on the
stellar position ($\alpha$, $\delta$). Here,
$(U_\odot,V_\odot,W_\odot)$ is the Sun velocity with respect to
the Local Standard of Rest (LSR), for which Dehnen \& Binney
(1998) estimated $(+10.00\pm 0.36,+5.25\pm 0.62, +7.17\pm 0.38)$
km s$^{-1}$ from the analysis of the Hipparcos catalogue. The
tangential velocities
 $V_\alpha$ and $V_\delta$ (km s$^{-1})$,
are computed from the observed proper motions (arcsec yr$^{-1}$)
and distances (pc) derived from trigonometric or photometric
parallaxes, $\pi=1/d$, as usual:
\begin{eqnarray}
 V_\alpha &=& 4.74047\,d\,\mu_{\alpha}\cos\delta \nonumber \\
 V_\delta &=& 4.74047\,d\,\mu_\delta  \nonumber
\end{eqnarray}

\subsection{Tangential velocity distribution}

If the full 3D space velocity cannot be recovered, as in the case
of proper motion surveys, we can adopt a similar procedure in the
2D tangential velocity plane, ($V_{\alpha}$, $V_{\delta}$). The
bivariate marginal distribution, $\psi(V_{\alpha},V_{\delta})$,
can be obtained by properly integrating the distribution in Eq.\
\ref{eq:UVWdistribution} along the $V_r$ component:



\begin{eqnarray}
\psi(V_{\alpha},V_{\delta}) &=& \frac{1}{2\pi\sigma_\alpha\sigma_\delta\sqrt{1-\rho^2}} \\ \nonumber
  && \exp\left[\frac{-1}{2(1-\rho^{2})} \left(\frac{(V_{\alpha}-V_{\alpha0})^{2}}{\sigma_{V\alpha}^2}
  - \right.\right. \\ \nonumber
  && \left.\left. 2\rho\frac{(V_{\alpha}-V_{\alpha0})}{\sigma_{V\alpha}} \frac{(V_{\delta}-V_{\delta0})}{\sigma_{V\delta}} +\frac{(V_{\delta}-V_{\delta0})^2}{\sigma_{V\delta}^2}\right)\right] \nonumber
 \label{eq:Valphadelta}
\end{eqnarray}
This is a general bivariate gaussian distribution which is defined
by five parameters: $V_{\alpha0}$, $V_{\delta0}$,
$\sigma_{V\alpha}$, $\sigma_{V\delta}$ and $\rho$.  These
parameters are linear functions of the first and second order
moments of Eq.\ \ref{eq:UVWdistribution}, as described for
instance in Trumpler \& Weaver (1953).

Our analysis will be based on  Eq.~3 
that represents the appropriate density distribution when radial
velocities are missing. \\ Notice that this approach, even in the
case of surveys involving widely different line-of-sights, allows
the derivation of the exact tangential velocity distribution for
every star, without any assumption on the unknown third velocity
component $V_{r}$.

\begin{figure}
\vspace{9cm} \includegraphics{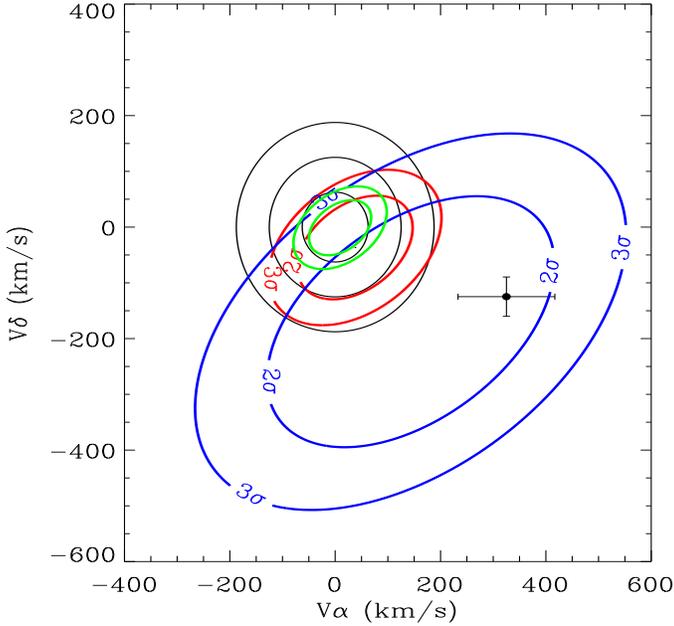}
  \caption{Tangential velocity distributions, $\psi(V_{\alpha},V_{\delta})$,
 toward the direction ($l=143^\circ$, $b=-57^\circ$) of WD0135-039 (solid circle).
 The ellipses show the iso-probability contours (2$\sigma$, 3$\sigma$) of
the thin disk, thick disk, and local halo populations based on the
kinematic parameters from Binney \& Merrifield (1998), Soubiran et
al.\ (2003) and Casertano, Ratnatunga \& Bahcall (1990),
respectively. The three concentric circles indicate the
 velocity thresholds, $V_{\rm min} = 4.74 \mu_{\rm lim} r $,
for the OHDHS survey ($\mu_{\rm lim}=0.33''$ yr$^{-1}$) at the
distances  $r=40$ pc, 80 pc and 120 pc. }
 \label{fig:t}
\end{figure}

\subsection{Thick disk model} \label{Sect:TDmodel}
The following properties for the population of thick disk WDs in
the solar neighborhood were assumed:
\begin{itemize}
\item a uniform local space density; for, the typical
 distance reachable by ground based surveys ($\sim$ 100 pc) is much smaller than
 the exponential vertical scale-height  of the thick disk ($h_z\simeq 1000$
 pc);
\item a velocity distribution (Eq.\ \ref{eq:UVWdistribution}) with
  ($\sigma_U, \sigma_V, \sigma_W, V_0) \simeq $ $(63,39,39,-45)$ km s$^{-1}$, as derived
  in Soubiran, Bienaym\`e \& Siebert (2003).
\end{itemize}

We notice that the velocity ellipsoid of the thick disk population
is not currently well established so that this choice will somehow
affect the final result. For instance, the presence of a
non-gaussian high velocity tail (cfr.\ Gilmore et al.\ 2002) would
increase the contamination affecting the halo WD sample.



\subsection{Kinematically selected samples}
In the case of a magnitude- {\it and} $\mu$-limited survey with a
total extension of $\Omega$ steradians, the following
observational constraints need to be taken into account:
\begin{enumerate}
 \item an apparent magnitude limit $m < m_{\rm lim}$ which implies a distance limit
as a function of the absolute magnitude, $M$, of the target:
$$ 
r < r_{\rm max}(M) = 10^{[0.2(m_{\rm lim} - M) + 1]}
$$ 
which, in turn, defines the maximum volume\footnote {Note that
this is a purely photometric definition which does not correspond
exactly to the analogue quantity adopted for the evaluation of the
WD density via the 1/$V_{\rm max}$ method (Schmidt 1975).},
$V_{\rm Max}(M)=\frac{1}{3}\Omega\, r_{\rm max}^3$ covered by the
survey;
 \item a proper motion limit $\mu > \mu_{\rm lim}$
which translates into a tangential velocity threshold varying with
stellar distance:
$$ 
 V_{\rm tan} = \sqrt{V_\alpha^2+ V_\delta^2} >  V_{\rm min}(r) = 4.74\mu_{\rm lim} r.
$$ 
\end{enumerate}

 Note that, although the distance distribution ($\propto r^2$) and the kinematic
 distribution (Eq.\ 3 
 ) of the complete
 population are independent, now they result correlated
 for the observed sample because of the existence of the velocity threshold,
 $V_{\rm min}(r)$.

The probability to select a star with absolute magnitude $M$ in
the range $(r,r+dr)$, ($V_{\alpha},V_{\alpha}+dV_{\alpha}$),
($V_{\delta},V_{\delta}+dV_{\delta}$) is then $dP =
f(r,V_{\alpha},V_{\delta})dr dV_{\alpha} dV_{\delta}$, where the
joint probability density is:
\begin{equation}
\label{Eq:f}
 f(r, V_\alpha, V_\delta) = \left\{
  \begin{array}{ll}
K r^2 \psi (V_\alpha, V_\delta) & \mbox{ if }  V_{\rm tan} > V_{\rm min}(r) \\
                                & \mbox{ and }   r < r_{\rm max}(M)\\
0  &  \mbox{ if }  V_{\rm tan} \le V_{\rm min}(r) \\
   &  \mbox{ or }   r \ge r_{\rm max}(M)\\
\end{array}
\right.
\end{equation}

Here, $K$ is a normalization constant such that $\int\int\int f\,
dr\,dV_\alpha\, dV_\delta=1$.


If we integrate over $r$ the joint probability density function
given in Eq.\ \ref{Eq:f}, we obtain the marginal density
distribution
\begin{eqnarray}
h(V_{\alpha},V_{\delta}) &=& \int_{0}^{r_{\rm max}}
f(r,V_{\alpha},V_{\delta})\,dr \label{eq:h}
\end{eqnarray}
which quantifies the probability that an object with tangential
velocities $(V_{\alpha},V_{\delta})$ could be randomly found
somewhere {\it within} the whole volume $ \frac{1}{3}\Omega\,
r_{\rm max}^3$, where an object with absolute magnitude $M$ could
in principle be observed.

At the same time, we can introduce the (conditional) probability
that an object with tangential velocities
$(V_{\alpha},V_{\delta})$ is found {\it at} the measured distance,
$r$:
 \begin{equation}
t(V_{\alpha},V_{\delta}|r) = \frac{f(r, V_\alpha, V_\delta) }{g(r) }
 \label{eq:t}
 \end{equation}
where
\begin{equation}
g(r) = \int\int_{-\infty}^{+\infty} f(r,V_{\alpha},V_{\delta})\,dV_{\alpha}\,dV_{\delta}
\end{equation}
is the marginal density distribution which defines the probability
that an object with whatever velocity can be observed at a
distance $r$.  Because the velocity threshold increases linearly
with distance, $V_{\rm min}\propto r$, the space distribution of
the proper-motion selected sample (Eq. 7) is also biased towards
smaller distances.

Both the marginal distribution $h(V_{\alpha},V_{\delta})$ and the
conditional probability $t(V_{\alpha},V_{\delta}|r)$ can be used
to test the consistency of each object with a parent population.
  In principle, the conditional probability
$t(V_{\alpha},V_{\delta}|r)$ seems more appropriate  than
$h(V_{\alpha},V_{\delta})$ since it fully utilizes the individual
stellar distances. However, the differences become insignificant
when the confidence level is set to sufficiently high values (see
next section).

 Note that, formally, Eq.\ \ref{eq:t} is equivalent to the
original distribution, $\psi(V_{\alpha},V_{\delta})$, except that
the probability is null for $\sqrt{V_\alpha^2+ V_\delta^2} \le
V_{\rm min}(r)$, and it has been re-normalized.

\subsection{Confidence intervals and contamination}


Basically, because a proper motion limited survey undersamples the
low velocity objects, the main difference between  the
kinematically selected distributions (Eqs.\ \ref{eq:h}-\ref{eq:t})
and the complete one (Eq.\ 3) 
 is that the probability density is redistributed from the low
velocity regions towards the high velocity tails.  This means that
the observed sample is biased towards high velocity objects, as
shown for instance by the simulations of Reyl\'e et al.\ (2001)
and Torres et al.\ (2002).
%
%

This effect needs to be taken into account when we define a
confidence interval over the $(V_{\alpha},V_{\delta})$ plane in
order to test the consistency with the parent population and to
estimate the contamination due to objects in the tails beyond the
critical limit.  In fact, the adoption of the original
$\psi(V_{\alpha},V_{\delta})$ to reject the disk stars with
respect to a certain confidence level, e.g.\ $1-\alpha=99\%$,
would exclude 99\% of {\it all} the existing thick disk stars
which, however, corresponds to a smaller fraction of the thick
disk objects that are really present in the kinematically selected
sub-sample. In this case, only the confidence interval defined for
$t(V_{\alpha},V_{\delta}|r)$, or $h(V_{\alpha},V_{\delta})$,
assures that the fraction of false negatives contaminating the
sample of {\it bona fide} halo stars does not exceed -- on average
-- 1\% of the {\it observed} thick disk objects.



\begin{figure*}
\vspace{8.5cm} \includegraphics{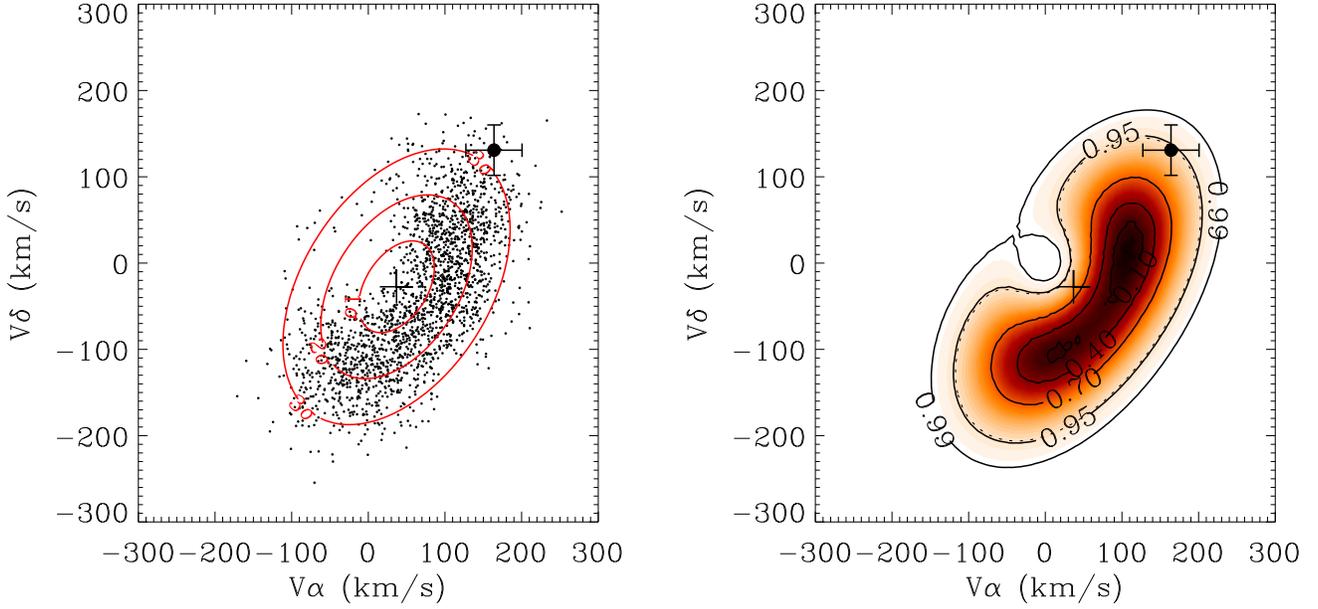}
 \caption{Left panel: iso-probability contours (1$\sigma$, 2$\sigma$, 3$\sigma$) of
$\psi(V_{\alpha},V_{\delta})$ compared against a Montecarlo
simulation (dots) of the thick disk stars in the direction
($l=226\fdg 34$,$b=-64\fdg 27$) of LHS 1447 (solid circle) drawn
from the kinematically selected distribution,
$h(V_{\alpha},V_{\delta})$. Right panel: iso-probability contours
(confidence levels of 40\%, 70\%, 95\% and 99\%) of
$h(V_{\alpha},V_{\delta})$.} \label{fig:h}
\end{figure*}

In the left panels of Figures \ref{fig:h}-\ref{fig:t} the
concentric ellipses show the
 iso-probability contours (1$\sigma$, 2$\sigma$, 3$\sigma$) of the
velocity distribution expected for thick disk stars,
$\psi(V_{\alpha},V_{\delta})$, evaluated in the direction of one
of the stars in the Oppenheimer's sample (LHS~1447), whose
tangential velocity is marked with a filled circle. The points
represent a Montecarlo realization of 2000 simulated WD's drawn
from the kinematically selected distributions
$h(V_{\alpha},V_{\delta})$ and $t(V_{\alpha},V_{\delta}|r)$. The
excess of ``simulated'' thick disk stars with high velocity is
evidenced by the fact that there are many more than $\sim$20
objects (1\% of the simulated sample) outside the $3\sigma$
confidence interval.


\begin{figure*}
\vspace{8.5cm} \includegraphics{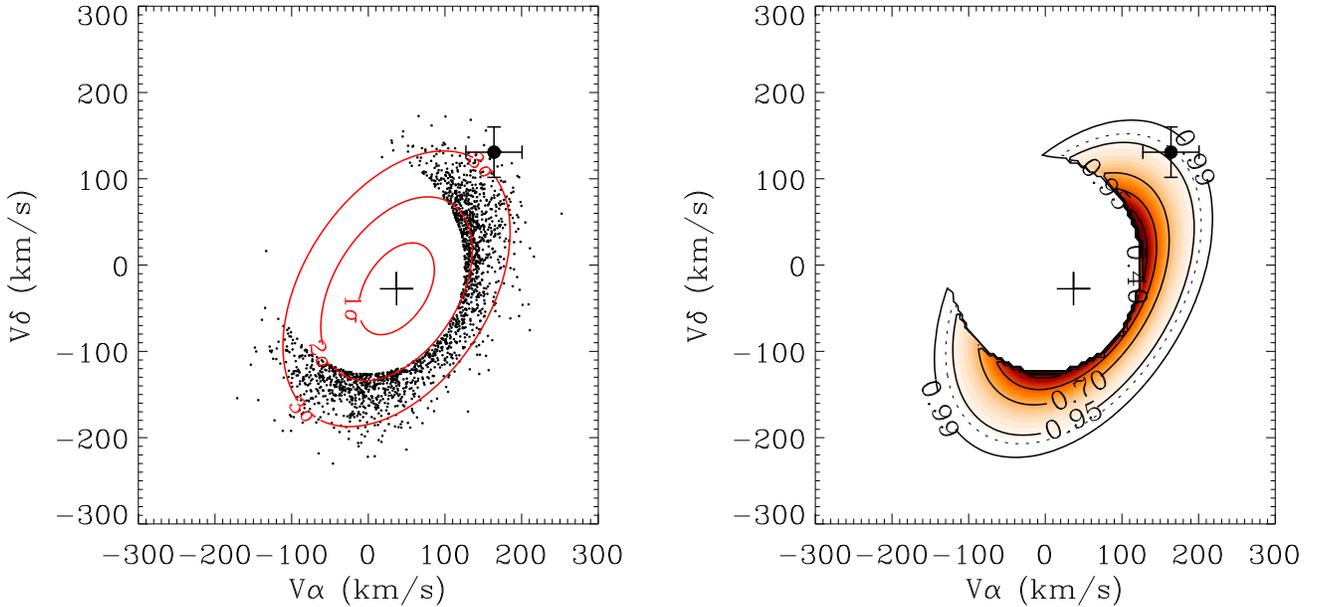}
  \caption{Left panel: iso-probability contours (1$\sigma$, 2$\sigma$, 3$\sigma$) of
$\psi(V_{\alpha},V_{\delta})$ compared against a Montecarlo
simulation (dots) of the thick disk stars in the direction ($l=
226\fdg 34$,$b=-64\fdg 27$) of LHS 1447 (solid circle) drawn from
the kinematically selected distribution,
$t(V_{\alpha},V_{\delta}|r)$. Right panel: iso-probability
contours (confidence levels of 40\%, 70\%, 95\% and 99\%) of
$t(V_{\alpha},V_{\delta}|r)$.}
 \label{fig:t}
\end{figure*}

LHS~1447 is also located outside the 3$\sigma$ contour so that,
according to the complete distribution, it should be rejected as a
thick disk star with a confidence level higher than
$1-\alpha=$99\%. Actually, that conclusion would be incorrect if
we tested the hypothesis that LHS 1447 is a member of the
kinematically selected sample as shown in the right panels of
Figures \ref{fig:h}-\ref{fig:t}, where the marginal and
conditional distributions, $h(V_{\alpha},V_{\delta})$ and
$t(V_{\alpha},V_{\delta}|r)$, are drawn. In fact, in these cases
the star is located {\it within} the iso-probability contour
delimiting the 99\% confidence level so that it must be accepted
as a thick disk star.




\section{Results and discussion}
Both distributions, $h(V_{\alpha},V_{\delta})$ and
$t(V_{\alpha},V_{\delta}|r)$, were used to analyze the WD sample
in the OHDHS survey.

The kinematic tests were carried out in the tangential plane of
each individual star so that no assumption on radial velocity is
necessary.
 The values of 95\% and 99\% for the confidence level ($1-\alpha$) were
chosen in order to minimize the presence of false negatives.
With a total sample of 98 WDs, presumably a mixture of (thin and
thick) disk and halo WDs, we expect that $<1$ (99\%) and $<5$
(95\%) of the high velocity thick disk stars would contaminate the
selected Pop.\ II WD sample.


\begin{table*}[t]
 \centering
 \caption{Estimation of the halo WD density based on the objects selected from the OHDHS sample.}

\begin{tabular}{rcccccl} \hline\hline

Confid. & \multicolumn{2}{c}{$\psi(V_{\alpha},V_{\delta})$} &
\multicolumn{2}{c}{$h(V_{\alpha},V_{\delta})$} &
\multicolumn{2}{c}{$t(V_{\alpha},V_{\delta}|r)$}\\
level & WDs & $\rho_{\rm WD}$ (M$_{\odot}$pc$^{-3}$) & WDs &
$\rho_{\rm WD}$
(M$_{\odot}$pc$^{-3}$) & WDs & $\rho_{WD}$ (M$_{\odot}$pc$^{-3}$) \\
 \hline
99\% & 14  & $(2.0\pm 0.9)\cdot10^{-5}$ & 10 & $(1.6\pm 0.8)\cdot10^{-5}$ & 10 & $(1.5\pm0.8)\cdot10^{-5}$\\
\hline
95\% & 20  & $(3.1\pm 1.0)\cdot10^{-5}$ & 12 & $(1.9\pm 0.9)\cdot10^{-5}$ & 13 & $(2.0 \pm0.9)\cdot10^{-5}$\\
\hline

\end{tabular}
 \label{tab:density}
\end{table*}

\begin{table*}
 \centering
 \caption{Estimation of the halo WD density based on the sample revised by  Salim et al.\ (2004).}

\begin{tabular}{rcccccl} \hline\hline

Confid. & \multicolumn{2}{c}{$\psi(V_{\alpha},V_{\delta})$} &
\multicolumn{2}{c}{$h(V_{\alpha},V_{\delta})$} &
\multicolumn{2}{c}{$t(V_{\alpha},V_{\delta}|r)$}\\
level & WDs & $\rho_{\rm WD}$ (M$_{\odot}$pc$^{-3}$) & WDs &
$\rho_{\rm WD}$
(M$_{\odot}$pc$^{-3}$) & WDs & $\rho_{\rm WD}$ (M$_{\odot}$pc$^{-3}$) \\
 \hline
99\% & 19  & $(1.8\pm 0.6)\cdot10^{-5}$ & 17 & $(1.6\pm 0.6)\cdot10^{-5}$ &  16 & $(1.5\pm0.6)\cdot10^{-5}$\\
\hline
95\% & 28  & $(3.3\pm 0.8)\cdot10^{-5}$ & 18 & $(1.8\pm 0.6)\cdot10^{-5}$ &  18 & $(1.9 \pm0.6)\cdot10^{-5}$\\
\hline

\end{tabular}
 \label{tab:density1}
\end{table*}

\subsection{Halo WD density}
In Table \ref{tab:density} we report the results based on this
procedure for the WD sample published by OHDHS.
 We only found 10 objects which do not appear consistent with the kinematically
selected density distributions, $h(V_{\alpha},V_{\delta})$ and
$t(V_{\alpha},V_{\delta}|r)$, at the 99\% confidence level, while
12-13 probable halo WDs are selected when $1-\alpha=95$\%.  As
expected, the number of candidates increases up to 14 (99\%) or 20
(95\%) in the case of a test based on the complete distribution,
$\psi(V_{\alpha},V_{\delta})$, mainly because of a higher
contamination.

Finally,  the halo WD density was estimated by means of the
classical 1/V$_{\rm Max}$ method (Schmidt 1975), and assuming a
value of 0.6 M$_\odot$ for the typical WD mass. The results, with
their (poissonian only) errors, are reported in Tab.\
\ref{tab:density}, where the different values refer to the two
confidence levels and the three probability distributions used for
the calculations.

Although affected by large uncertainties, the values in Tab.\
\ref{tab:density} suggest a density of $\rho_{\rm WD}\approx$
$10^{-5}$ M$_{\odot}$pc$^{-3}$, i.e.\ 0.1-0.2\% of the local dark
matter, which is an order of magnitude smaller than what reported
in OHDHS.

Our results are consistent with the local mass density of halo WDs
estimated by Gould et al.\ (1998), and with various reanalyses of
the OHDHS sample (e.g.\ Reid et al.\ 2001, Reyl\'e et al.\ 2001,
Torres et al.\ 2002, Salim et al.\ 2004). Furthermore, Carollo et
al.\ (2004), applying the statistical methodology described in
this paper on a new high proper motion survey based on GSC-II
material, derived a similar value of $\sim 10^{-5}$
M$_{\odot}$pc$^{-3}$.

\medskip

 Lacking
individual trigonometric parallaxes, a critical point of this (and
any) analysis is the choice of the method for the estimation of
the distances, which directly affects the evaluation of the WD
tangential velocities and, of course, of their stellar density. As
remarked by several authors (see e.g.\ Torres et al.\ 2002,
Bergeron 2003), empirical and theoretical CM relations can both
give rise to systematic errors.

To this regard, if for the distances of the OHDHS sample we adopt
the values recently redetermined\footnote{They adopted CM
relations based on theoretical cooling tracks of 0.6 M$_{\odot}$
WDs with H or He atmospheres. This resulted in distances 16\%
systematically larger (on average) than those in OHDHS.} by Salim
et al.\ (2004), the number of selected halo WDs increases but the
resulting densities, shown in Table \ref{tab:density1}, are not
significantly different from those reported in Table
\ref{tab:density}.

\subsection{Distance and velocity errors}
The large error, $\sim$ 20-30\%, affecting WD photometric
parallaxes, cannot be neglected in a rigorous statistical
analysis. Basically, besides the contribution of the photometric
errors, the large uncertainty in the distance modulus, $m-M$,
derives from the large {\it intrinsic} dispersion ($\sigma_{\rm
Mv}\simeq 0.4$ - 0.5 mag) of the CM relation, a consequence of the
superposition of cooling sequences of WDs of different masses and
atmospheres.


\begin{table*}[t]
 \centering
 \caption{Same as Tab.\ref{tab:density} after adopting a thick disk velocity
 distribution convolved with the observation errors.  }

\begin{tabular}{rcccccl} \hline\hline

Confid. & \multicolumn{2}{c}{$\psi(V_{\alpha},V_{\delta})$} &
\multicolumn{2}{c}{$h(V_{\alpha},V_{\delta})$} &
\multicolumn{2}{c}{$t(V_{\alpha},V_{\delta}|r)$}\\
level & WDs & $\rho_{\rm WD}$ (M$_{\odot}$pc$^{-3}$) & WDs &
$\rho_{\rm WD}$
(M$_{\odot}$pc$^{-3}$) & WDs & $\rho_{\rm WD}$ (M$_{\odot}$pc$^{-3}$) \\

\hline
99\% &  6  & $(1.3\pm 0.8)\cdot10^{-5}$ &  3 & $(1.2\pm 0.8)\cdot10^{-5}$ & 3 & $(1.2\pm0.8)\cdot10^{-5}$\\
\hline
95\% &  14  & $(2.0\pm 0.9)\cdot10^{-5}$ & 5 & $(1.3\pm 0.8)\cdot10^{-5}$ & 6 & $(1.3 \pm0.8)\cdot10^{-5}$\\
\hline

\end{tabular}
 \label{tab:convdensity}
\end{table*}

\begin{table*}
 \centering
 \caption{Same as Tab.\ref{tab:density1} after adopting a thick disk velocity
 distribution convolved with the observation errors.  }

\begin{tabular}{rcccccl} \hline\hline

Confid. & \multicolumn{2}{c}{$\psi(V_{\alpha},V_{\delta})$} &
\multicolumn{2}{c}{$h(V_{\alpha},V_{\delta})$} &
\multicolumn{2}{c}{$t(V_{\alpha},V_{\delta}|r)$}\\
level & WDs & $\rho_{\rm WD}$ (M$_{\odot}$pc$^{-3}$) & WDs &
$\rho_{\rm WD}$
(M$_{\odot}$pc$^{-3}$) & WDs & $\rho_{\rm WD}$ (M$_{\odot}$pc$^{-3}$) \\

\hline
99\% &  14  & $(1.2\pm 0.5)\cdot10^{-5}$ &  8 & $(0.7\pm 0.4)\cdot10^{-5}$ &  5 & $(0.6\pm0.4)\cdot10^{-5}$\\
\hline
95\% &  18  & $(1.7\pm 0.6)\cdot10^{-5}$ &  14 & $(1.2\pm 0.5)\cdot10^{-5}$ & 9 & $(1.0 \pm0.5)\cdot10^{-5}$\\
\hline

\end{tabular}
 \label{tab:convdensity1}
\end{table*}

In practice, the main effect of the tangential velocity errors,
$\epsilon_V/V = \sqrt{(\sigma_\mu/\mu)^2+(\sigma_d/d)^2}$, is to
increase the dispersion and the overlap of the ``observed''
kinematic distributions belonging to the various stellar
populations. Clearly this also increases the contamination of the
disk  WDs and makes the identification of the halo WDs more
difficult.

Although a more rigorous statistical analysis should be necessary
to consider properly the presence of these errors, a conservative
estimation can be given by selecting only those objects which are
not consistent with the ``observed'' kinematic distribution that
results from convolving the
 projected kinematic distribution of the thick disk
(Eq.~3) with a bivariate gaussian error distribution with null
mean and dispersions, ($\epsilon_{V\alpha},
\epsilon_{V\delta})_{(i)}$, corresponding to the velocity errors
of the $i$-th object. The velocity errors have been derived by
assuming the proper motion errors, $\sigma_\mu$, listed in Tab.\ 1
of OHDHS, and a more realistic photometric parallax error,
$\sigma_d/d$, of 25\% (instead of 20\%).


The different halo WD densities estimated from the objects which
are not consistent (at the 95\% and 99\% confidence level) with
the new distributions are reported in Tables 3 and 4. Because of
the larger velocity thresholds, the number of selected halo WDs is
smaller than those reported in Tables 1 and 2. The estimated WD
densities, uncorrected for the loss of halo WDs with disk
kinematics, decrease proportionally, but are still consistent with
$\rho_{\rm WD} \sim 10^{-5}$ M$_{\odot}$pc$^{-3}$. Note that the
minimum values, which are reported in Tab. 4, have been derived
from the data of Salim et al. (2004) who provided distances (and
thus volumes) systematically larger than OHDHS.
\\

%
%
%
%
%
%
%


\subsection{On the thick disk model}
As mentioned in Sect.\ \ref{Sect:TDmodel},
 our selection criterion depends implicitly
 also on the choice of the kinematic parameters
adopted for the thick disk, whose spatial and kinematical
properties are still matter of debate and investigation. Here, we
have used the velocity ellipsoid recently derived by Soubiran et
al.\ (2003) from a sample of $\sim$ 400 giants with 3D kinematics
at a distance of 200-800 pc towards the North Galactic Cap. Their
results are very close to the kinematic parameters
estimated\footnote{Casertano, Ratnatunga \& Bahcall (1990) derived
($\sigma_U, \sigma_V, \sigma_W, V_0) \simeq $ $(66,37,38,-40)\pm
10$ km s$^{-1}$ from a maximum likelihood analysis of high proper
motion stars within 500 pc of the Sun. } by Casertano, Ratnatunga
\& Bahcall (1990) and are consistent with various other
determinations of the thick disk kinematics, which support
velocity dispersions of 40-60 km s$^{-1}$ and an asymmetric drift
in the range 30-50 km s$^{-1}$.

 Although controversial, some authors claim the presence
of a vertical velocity gradient, that supports a thick disk which
rotates faster close to the galactic plane (i.e.\ where the WD
sample is localized), than at higher Z's, where the studies of the
thick disk kinematics have been usually carried out.
 In particular, Chiba \& Beers (2000), who analyzed
1203 metal poor stars non-kinematically selected, found a rapidly
rotating thick disk close to the galactic plane with a small
asymmetric drift $V_0\simeq -20$ km s$^{-1}$ and
with velocity dispersions ($\sigma_U, \sigma_V, \sigma_W) \simeq $
$(46\pm 4,50\pm 4, 35\pm 3)$ km s$^{-1}$. Moreover, they
 determined a velocity gradient  $\partial V_0/\partial |Z| \simeq
-30\pm 3$ km s$^{-1}$ kpc$^{-1}$, that, however, other studies
(e.g.\ Soubiran et al. 2003) do not detect.
 Nevertheless, a fast rotating thick disk at $Z\approx 0$ was
determined\footnote{ They
 estimated a rotation lag of $V_0\simeq -28.3\pm 3.8$ km
s$^{-1}$ for the ``old'' disk component  with dispersions
($\sigma_U, \sigma_V, \sigma_W) \simeq $ $(56.1\pm 3.9,34.2\pm
2.5,31.2\pm 2.5)$ km s$^{-1}$.}
 also by Upgren et al.\ (1997) from a sample of
K-M dwarfs in the solar neighborhood ($d \la 50$ pc) with
trigonometric parallaxes and proper motions
 from the Hipparcos catalogue and radial velocity measurements.

Thus, in order to test the sensitivity of our method with respect
to the adopted thick disk model, we repeated the WD selection of
the Salim et al.\ (2003) sample through the distributions
$h(V_{\alpha},V_{\delta})$ and $t(V_{\alpha},V_{\delta}|r)$
derived using the velocity ellipsoid from Chiba \& Beers (2000).
The new results are consistent (within 1$\sigma$) with the values
obtained with the kinematics from Soubiran et al.\ (2003),
although the resulting densities appear typically larger than the
previous ones.

 For instance, with a 99\% confidence level
 we find $\rho_{\rm WD} \simeq (1.7\pm 0.6) 10^{-5}$
M$_{\odot}$pc$^{-3}$ for both $h(V_{\alpha},V_{\delta})$ and
$t(V_{\alpha},V_{\delta}|r)$ when the velocity errors are not
taken into account (cfr.\ Tab.\ \ref{tab:density1}), while the
distributions convolved with the velocity errors provide
$\rho_{\rm WD} \simeq (0.9\pm 0.5) 10^{-5}$ M$_{\odot}$pc$^{-3}$
 (cfr.\ Tab.\ \ref{tab:convdensity1}).
 The 95\% confidence level also provides similar but
 systematically higher new densities up to  $(1.8\pm 0.7) 10^{-5}$
M$_{\odot}$pc$^{-3}$ and $(3.0\pm 0.9) 10^{-5}$
M$_{\odot}$pc$^{-3}$ respectively when the velocity errors are, or
are not, convolved with the tangential velocity  distributions.

Anyhow, it appears that, with the adopted confidence levels,
significantly higher density (e.g.\ close to $\sim 10^{-4}$
M$_{\odot}$pc$^{-3}$ may be attained only with disk ellipsoids
kinematically much ``cooler'' than those expected for a typical
thick disk population.
 For instance, a total density  $(8.8\pm 0.2) 10^{-5}$
M$_{\odot}$pc$^{-3}$ is only obtained counting all the 41 WDs
which are not consistent with the {\it thin} disk\footnote {We
adopted ($\sigma_U, \sigma_V, \sigma_W, V_0) \simeq $ $(34, 21,
18; -6)$ km s$^{-1}$ from Tab. 10.4 of Binney \& Merrifield
(1998).} kinematics (using $t(V_{\alpha},V_{\delta}|r)$ with a
95\% confidence level), i.e.\ summing both halo {\it and} thick
disk WDs.

\begin{figure*}
\vspace{6.5cm} \includegraphics{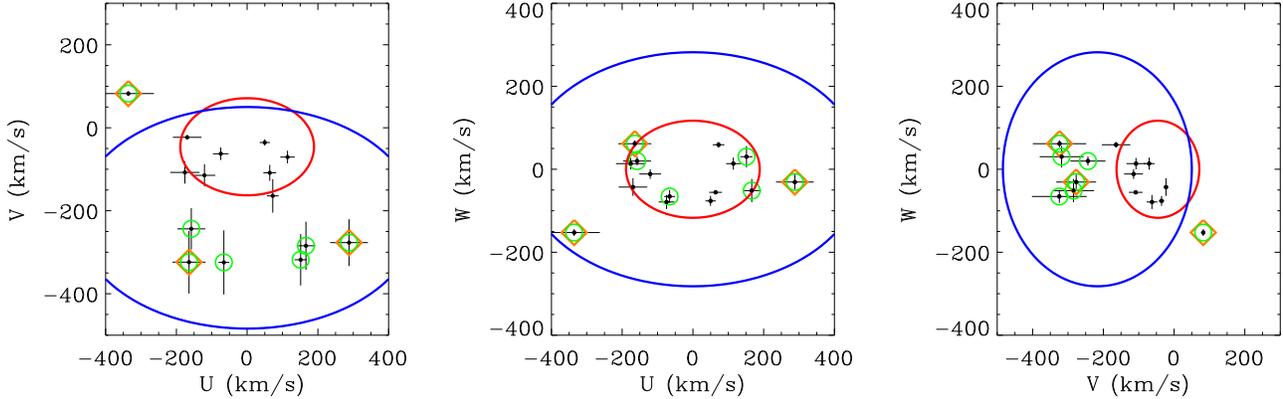}
 \caption{Velocity distribution (U,V,W) of the subsample of 15 stars with available radial
 velocity (dots with 1$\sigma$ error bars) from Salim et al.\ (2004).
 The objects selected by means of the
 distributions $h(V_{\alpha},V_{\delta})$ and
$t(V_{\alpha},V_{\delta}|r)$ with a 95\% confidence level (Tab.\
\ref{tab:convdensity1}) are marked with circle and diamond
symbols, respectively. The ellipses show the $3\sigma$
iso-probability contours of the thick disk and halo velocity
distribution.} \label{fig:UVW}
\end{figure*}

\subsection{UVW distribution}
 Salim et al.\ (2003) provide radial velocities for 15 DA WDs, 13
of which derived from new measurements of the OHDHS sample and two
from Pauli et al.\ (2003), so that, in principle, a more accurate
kinematic membership for these objects may be inferred using the
information from the full 3D velocities.
 This requires 3D velocity distributions for kinematically selected
samples which are beyond the scope of the current study.
 However, the availability of both tangential and radial
 velocities
 for this subsample offers the possibility to check {\it a posteriori}
 the efficiency of the 2D kinematic analysis  adopted in this work
 and described in Sect.\ \ref{Sect:kinematics}.

  To this regard, Figure \ref{UVW} shows the
(U,V,W) velocities derived from Eq.\ \ref{UVW} for the 15 stars
with available radial velocity.
 Those which have been selected with a 95\% confidence level by means of the
 distributions $h(V_{\alpha},V_{\delta})$ and $t(V_{\alpha},V_{\delta}|r)$ convolved
 with the velocity errors
(Tab.\ \ref{tab:convdensity1}) are marked with square and diamond
symbols. In addition, the $3\sigma$ iso-probability ellipses of
the thick disk and halo velocity distributions, based on the
kinematic parameters respectively from Soubiran et al.\ (2003) and
Casertano, Ratnatunga \& Bahcall (1990), are also plotted.
 The three panels of Fig.\ \ref{UVW} indicate that, basically, all
the likely halo WDs have been properly identified by our kinematic
analysis based on the 2D $(V_{\alpha},V_{\delta})$ distributions,
thus supporting the reliability of our selection procedure.



\section{Conclusions}
A kinematically selected sample made of 98 WDs with $\mu>$
0.33$\arcsec yr^{-1}$ was published by OHDHS who performed a high
proper motion survey over 4165 deg$^2$ toward the SGP down to
$R59F \simeq 19.8$.  These data stimulated a number of studies
addressing the issue that a significant part of the dark halo of
the Milky Way could be composed of matter in the form of ancient
cool WDs. The basic problem -- as addressed by several authors --
is the criterion to disentangle the mixture of (thick) disk and
halo objects on the basis of their kinematic properties and ages.

To this regard, we have implemented a general method for the
kinematic analysis of high proper motion surveys and applied it to
the identification of reliable halo stars.
 The kinematically-selected tangential velocity distributions are
derived for every star, so that no assumption on the unknown third
velocity component, $V_{r}$, nor any approximation on the galactic
components (U,V,W), is necessary.

 We selected  as {\it
bona fide} halo WDs only those stars whose tangential velocity is
inconsistent, at the 95\% and 99\% confidence levels, with the
appropriate projected distribution, $h(V_{\alpha},V_{\delta})$ or
$t(V_{\alpha},V_{\delta}|r)$, of the observed thick disk
population, thus assuring limited contamination of thick disk
objects.
%
Finally, the effect of large velocity errors, which derive from
the intrinsic uncertainty of the WD photometric parallaxes, was
also discussed and taken into account.

We applied this methodology to the OHDHS sample and selected 10
probable halo WDs (that became 3 after the inclusion of the
velocity errors) at the a 99\% confidence level. Through the
1/V$_{\rm Max}$ method, we estimated  a local WD density of
$\rho_{\rm WD}\simeq 1 \div 2 \cdot 10^{-5}$ M$_{\odot}$pc$^{-3}$
(i.e.\ 0.1-0.2\% of the local dark matter) which is consistent
with the values found by Gould et al.\ (1998), as well as by other
authors who reanalyzed the OHDHS sample (e.g.\ Reid et al.\ 2001,
Reyl\'e et al.\ 2001, Torres et al.\ 2002, Flynn et al.\ 2003).
 The same methodology applied to the OHDHS sample
revised by Salim et al.\ (2004) yields a similar value. These
results agree with those found by Carollo et al.\ (2004) from a
first analysis of new data of an independent high proper motion
survey in the Northern hemisphere based on material and procedures
used for the construction of the GSC-II.

Although affected by a large uncertainty due to the small
statistics and low accuracy of the photometric parallaxes, our
results clearly indicate that ancient cool WDs do {\it not}
contribute significantly to the baryonic fraction of the galactic
dark halo, as possibly suggested by the microlensing experiments
which claimed that $\sim$ 20\% of the dark matter is formed by
compact objects of $\sim$ 0.5 M$_{\odot}$ (Alcock et al. 2000).

\begin{acknowledgements}
We wish to acknowledge the useful discussions with R.\ Drimmel,
S.T.\ Hodgkin, B.\ McLean, R. Smart, and L.\ Terranegra. We would
also like to thank the anonymous referee for valuable comments on
the submitted manuscript.

Partial financial support to this research came from the Italian
Ministry of Research (MIUR) through the COFIN-2001 program.

\end{acknowledgements}

\end{document}